\documentclass[12pt,a4paper]{article}
\usepackage{amsmath,amsfonts,euscript,amssymb}
\usepackage{graphicx}

\newcommand{\be}{\begin{equation}}
\newcommand{\ee}{\end{equation}}
\newcommand{\bea}{\begin{eqnarray}}
\newcommand{\eea}{\end{eqnarray}}

\begin{document}
\textwidth=135mm
\textwidth=150mm
\textheight=200mm
\begin{center}
{\bfseries INTRINSIC TIME IN FRIEDMANN -- ROBERTSON -- WALKER UNIVERSE}
\vskip 5mm
Alexander Pavlov\\
Bogoliubov~Laboratory~for~Theoretical~Physics,
Joint~Institute~of~Nuclear~Research,
~Joliot-Curie~street~6,~Dubna,~141980,~Russia; \\
Institute of Mechanics and Energetics\\
Russian State Agrarian University --\\
Moscow Timiryazev Agricultural Academy, Moscow, 127550, Russia\\
alexpavlov60@mail.ru
\end{center}




\section{Global intrinsic time in FRW universe}

In Geometrodynamics there is a many-fingered intrinsic time as a scalar field \cite{PP, Pavlovpreprint, PavlovpreprintExact}.
A global time exists in homogeneous cosmological models (see, for example, papers
\cite{Misner,Kasner,Kuchar,Pavlov,PavlovFlat,Pavlov1}).
The observational Universe with high precision is homogeneous and isotropic \cite{Planck}.
So, it is natural to choose a compact conformally flat space of modern scale $a_0$
as a background space. It is necessary to study an applicability of the intrinsic global time chosen to nearest non-symmetric cases by
taking into account linear metric perturbations. 
The first quadratic form
$${\bf f}=f_{ij}({\bf x})dx^i\otimes dx^j$$
in spherical coordinates  $\chi, \theta, \varphi$
\begin{equation}\label{back}
a_0^2\left[d\chi^2+\sin^2\chi (d\theta^2+\sin^2\theta d\varphi^2)\right]\equiv
a_0^2({}^1f_{ij})dx^i dx^j
\end{equation}
should be chosen for a background space of positive curvature.
We introduced the metric of invariant space $({}^1f_{ij})$ with a curvature ${\cal K}=\pm 1, 0$.
For a pseudosphere in (\ref{back}) instead of $\sin\chi$ one should take $\sinh\chi$, and for a flat space
one should take $\chi.$ The spatial metric is presented as
\begin{equation}\label{Fmetric}
(\gamma_{ij})=a^2(t)({}^1f_{ij})=e^{-2D}(f_{ij})=\left(\frac{a}{a_0}\right)^2\tilde\gamma_{ij}.
\end{equation}
Hence, one yields an equality:
$\tilde\gamma_{ij}=f_{ij},$
so, in our high symmetric case, the conformal metric is equal to the background one.

The energy-momentum tensor is of the form
$$
(T_{\mu\nu})=\left(
\begin{array}{cc}
N^2 \rho&0\\
0&p\gamma_{ij}
\end{array}
\right),
$$
where $\rho$ is a background energy density of matter, and $p$ is its pressure.
We have for the matter contribution
\begin{equation}\label{matterT}
\sqrt\gamma T_{\bot\bot}=\sqrt{({}^1 f)}a^3\rho ,
\end{equation}
because of
$$T_{\bot\bot}=n^\mu n^\nu T_{\mu\nu}=\frac{1}{N^2}T_{00}=
\rho,\qquad n^\mu=\left(\frac{1}{N},-\frac{N^i}{N}\right).$$

The functional of action in the symmetric case considered here takes the form
\begin{equation}\label{Wh}
W^{(0)}=-\int\limits_{t_I}^{t_0}\, dt\int\limits_{\Sigma_t}\, d^3x
\pi_D\frac{d D}{dt}-\int\limits_{t_I}^{t_0}\, dt\int\limits_{\Sigma_t}\, d^3x
N{\cal H}_\bot .
\end{equation}
Let us present results of necessary calculations been executed
\begin{eqnarray}
K_{ij}&=&-\frac{1}{2N}\frac{d\gamma_{ij}}{dt}=\frac{1}{N}\frac{d D}{dt}\gamma_{ij},\nonumber\\
K&=&\gamma^{ij}K_{ij}=\frac{3}{N}\frac{d D}{dt},\nonumber\\
\left(K_{ij}K^{ij}-K^2\right)&=&-\frac{6}{N^2}\left(\frac{d D}{dt}\right)^2,\nonumber\\
\pi_D\frac{d D}{dt}&=&
4\sqrt\gamma K\frac{d D}{dt}=\frac{12}{N}\sqrt{f}e^{-3D}\left(\frac{d D}{dt}\right)^2,\nonumber\\
R&=&\frac{6 {\cal K}}{a_0^2}e^{2D}=\frac{6{\cal K}}{a^2}.\nonumber
\end{eqnarray}
The Hamiltonian constraint
\begin{equation}\label{Ham0Class}
{\cal H}_\bot=\sqrt{\gamma}\left(K_{ij}K^{ij}-K^2\right)-\sqrt{\gamma}R+\sqrt{\gamma} T_{\bot\bot},
\end{equation}
takes the following form
$$
{\cal H}_\bot=-6\sqrt{f}e^{-3D}\left[\frac{1}{N^2}
\left(\frac{dD}{dt}\right)^2+\frac{\cal K}{a_0^2}e^{2D}-\frac{1}{6}\rho \right].
$$

The action (\ref{Wh}) presents a model of the classical mechanics \cite{PP}
\begin{equation}\label{Whint}
W^{(0)}=V_0\int\limits_{t_I}^{t_0}\, dt
\left[-\frac{6}{N}e^{-3D}\left(\frac{d D}{dt}\right)^2+\frac{6{\cal K}}{a_0^2}N e^{-D}-
N e^{-3D}\rho \right]
\end{equation}
after integration over the slice $\Sigma_0$, where
\begin{equation}\label{Sigma0}
V_0:=\int\limits_{\Sigma_0}\sqrt{f}\, d^3x
\end{equation}
is a volume of the space with a scale $a_0$. We rewrite it in Hamiltonian form
\begin{equation}\label{LG}
W^{(0)}=\int\limits_{t_I}^{t_0}\, dt\left[p_D\frac{d D}{dt}-N H_{\bot}\right],
\end{equation}
where
\begin{equation}\label{clHam}
H_{\bot}=-\frac{1}{24V_0}e^{3D}p_D^2-\frac{6{\cal K}V_0}{a_0^2}e^{-D}+V_0e^{-3D}\rho
\end{equation}
is the Hamiltonian constraint as in classical mechanics.
Here $D(t)$ is a generalized coordinate, and $p_D$ is its canonically conjugated momentum
\begin{equation}\label{pa}
p_D=\frac{\delta W^{(0)}}{\delta\dot D} =-\frac{12 V_0}{N}e^{-3D}\left(\frac{d D}{dt}\right).
\end{equation}
Resolving the constraint (\ref{clHam}), we get the energy
\begin{equation}\label{paE}
p_D^2=E^2 (D),\qquad E(D):= 2\sqrt{6}V_0 e^{-2D}\sqrt{e^{-2D}\rho -\frac{6{\cal K}}{a_0^2}},
\end{equation}
that was lost in Standard cosmology \cite{MY}.

The equation of motion is obtained from the action (\ref{LG}):
\begin{equation}\label{eqnmotion}
\dot{p}_D=N\left[
\frac{e^{3D}}{8V_0}p_D^2-\frac{6{\cal K}V_0}{a_0^2}e^{-D}+3V_0e^{-3D}\rho-V_0e^{-3D}
\frac{\partial{\rho}}{\partial D}
\right] .
\end{equation}

From the Hamiltonian constraint (\ref{clHam}), the equation of motion (\ref{eqnmotion}), and the energy
continuity equation
\begin{equation}\label{eceqn}
\dot\rho=-3(\rho+p)\frac{\dot{a}}{a},
\end{equation}
we get Einstein's equations in standard form:
\begin{eqnarray}
\left(\frac{\dot{a}}{a}\right)^2+\frac{\cal{K}}{a^2}&=&\frac{1}{6}\rho,\nonumber\\
2\frac{\ddot{a}}{a}+\left(\frac{\dot{a}}{a}\right)^2+\frac{\cal{K}}{a^2}&=&-\frac{1}{2}p.\nonumber
\end{eqnarray}
Here we have used the normal slicing condition $N=1$.

Substituting $p_D$ from (\ref{pa}) into (\ref{paE}), we obtain the famous {\it Friedmann equation}
\begin{equation}\label{Friedmann}
\frac{1}{N^2 e^{2D}}{\left(\frac{d D}{dt}\right)}^2+\frac{\cal K}{a_0^2}=\frac{1}{6}e^{-2D}\rho .
\end{equation}
From the Standard cosmology point of view, it connects the expansion rate of the Universe (Hubble parameter)
\begin{equation}\label{Hubblep}
H:=\left(\frac{\dot{a}}{a}\right)=-\frac{d D}{dt}
\end{equation}
with an energy density of matter $\rho$ and a spatial curvature
\begin{equation}\label{Hubble0}
{H^2\equiv \left(\frac{d D}{dt}\right)^2= \frac{1}{6}
\left(\rho_M+\rho_{rigid}+\rho_{rad}+\rho_{curv}\right).}
\end{equation}
In the right side of equation (\ref{Hubble0}),
$\rho_M$ is an energy density of nonrelativistic matter
$$\rho_M=\rho_{M,0}\left(\frac{a_0}{a}\right)^3,$$
$\rho_{rigid}$ is an energy density of matter
$$\rho_{rigid}=\rho_{rigid,0}\left(\frac{a_0}{a}\right)^6,$$
with a rigid state equation \cite{Zel'}
$${p=\rho,}$$
$\rho_{rad}$ is an energy density of radiation
$$\rho_{rad}=\rho_{rad,0}\left(\frac{a_0}{a}\right)^4,$$
$\rho_{curv}$, which is defined as
$$\frac{1}{6}\rho_{curv}:= -\frac{\cal K}{a_0^2}e^{2D}=-\frac{\cal K}{a^2},$$
is a contribution from the spatial curvature. In the above formulae
$\rho_{M,0};$ $\rho_{rigid,0};$ $\rho_{rad,0};$ $\rho_{curv,0}$ are modern values of the densities.
Subsequently, the equation for the energy (\ref{paE}) can be
rewritten as
$$E(D)=2\sqrt{6}V_0e^{-3D}\sqrt{\rho+\rho_{curv}}.$$
The $CDM$ model considered has not dynamical degrees of freedom.
{\it According to the Conformal cosmology interpretation,
the Friedmann equation (\ref{Hubble0}) has a following sense:
it ties the intrinsic time interval $dD$ with the coordinate time
interval $dt$, or the conformal interval $d\eta=dt/a$.}
Let us note, that in the left side of the Friedmann equation (\ref{Hubble0}) we see the square of the
extrinsic York's time which is proportional to the square of the Hubble parameter. If we choose the
extrinsic time, the equation (\ref{Hubble0}) becomes algebraic of the second order,
and the connection between temporal intervals should be
lost\footnote{``The time is out of joint''. William Shakespeare: {\it Hamlet}. Act 1. Scene V. Longman,
London (1970).}.

In an observational cosmology, the density can be expressed in terms of the present-day
critical density $\rho_{c}$:
$$
\rho_c (a)\equiv 6H_0^2,
$$
where
$H_0$ is a modern value of the Hubble parameter. Further, it is convenient to use density parameters as ratio
of present-day densities
$$
\Omega_{M}\equiv\frac{\rho_{M,0}}{\rho_c},\quad \Omega_{rigid}\equiv\frac{\rho_{rigid,0}}{\rho_c},\quad
\Omega_{rad}\equiv\frac{\rho_{rad,0}}{\rho_c},\quad
\Omega_{curv}\equiv\frac{\rho_{curv,0}}{\rho_c},
$$
satisfying the condition
$$
\Omega_{M}+\Omega_{rigid}+\Omega_{rad}+\Omega_{curv}=1.
$$
Then, one yields
\begin{eqnarray}\nonumber
H^2&=&\frac{1}{6}\rho_c\left[\Omega_{M}\left(\frac{a_0}{a}\right)^3+\Omega_{rigid}
\left(\frac{a_0}{a}\right)^6+
\Omega_{rad}\left(\frac{a_0}{a}\right)^4+
\Omega_{curv}\left(\frac{a_0}{a}\right)^2\right].\nonumber
\end{eqnarray}
According to NASA diagram, 25\% of the Universe is dark matter,
70\% of the Universe is dark energy about which practically nothing is known.

After transition to conformal variables
\begin{equation}\label{confvar}
Ndt=a_0e^{-D}d\eta,\qquad \tilde\rho=e^{-4D}\rho ,
\end{equation}
the Friedmann equation takes the form
\begin{equation}
\left(\frac{d D}{d\eta}\right)^2+{\cal K}=\frac{1}{6}a_0 N^2 e^{2D}\tilde\rho.
\end{equation}

\section{Global time in perturbed FRW universe}

Let us consider additional corrections to the Friedmann equation from non ideal FRW model, taking into account
metric perturbations.
The metric of perturbed FRW universe can be  presented as
\begin{equation}\label{gmnperturb}
g_{\mu\nu}=a^2(t)\left({}^1 f_{\mu\nu}+h_{\mu\nu}\right),
\end{equation}
where $({}^1 f_{\mu\nu})$ is the metric of spacetime with the spatial metric $({}^1 f_{ij})$ considered above,
deviations $h_{\mu\nu}$ are assumed small.
The perturbation $h_{\mu\nu}$ is not a tensor in the perturbed universe, nonetheless we define
$$
h_\mu^\nu\equiv ({}^1 f^{\nu\rho})h_{\rho\mu},\qquad
h^{\mu\nu}\equiv({}^1 f^{\mu\rho})({}^1 f^{\nu\sigma})h_{\rho\sigma}.
$$

For a coordinate system chosen in the background space, there are various possible coordinate systems in the
perturbed spacetime. In GR perturbation theory, {\it a gauge transformation} means a coordinate transformation
between coordinate systems in the perturbed spacetime.  The coordinates of the background spacetime are kept
fixed, the correspondence between the points in the background and perturbed spacetime is changing.
A manifestly gauge invariant cosmological perturbation theory was built by
James Bardeen \cite{Bardeen}, and analyzed in details by Hideo Kodama and Misao Sasaki \cite{Kodama}.
Now, keeping the gauge chosen, {\it id est} the correspondence between the background and perturbed spacetime
points, we implement coordinate transformation in the background spacetime. Because of our background coordinate
system was chosen to respect the symmetries of the background, we do not want to change our slicing.
Eugene Lifshitz made decomposition of perturbations of metric and energy -- momentum tensor into scalar, vortex,
and tensor contributions refer to their transformation properties under rotations in the background space
in his pioneer paper \cite{Lifshitz}.
The scalar perturbations couple to density and pressure perturbations. Vector perturbations couple to rotational
velocity perturbations. Tensor perturbations are nothing but gravitational waves.

$\bullet$
In a flat case (${\cal K}=0$) the eigenfunctions of the Laplace -- Beltrami operator are plane waves.
For arbitrary perturbation $f(t, {\bf x})$ we can make an expansion
$$f(t, {\bf x})=\sum\limits_{{\bf k}=0}^\infty f_{\bf k}(t)e^{\imath {\bf k}\cdot{\bf x}}$$
over Fourier modes. We can consider them in future as a particular case of models with constant curvature.
Let us proceed the harmonic analysis of linear geometric perturbations using irreducible representations of
isometry group of the corresponding constant curvature space \cite{Durrer,Ruth}.

$\bullet$ Scalar harmonic functions.

Let us define for a space of positive curvature ${\cal K}=1$ an invariant space with the first quadratic form
\begin{equation}\label{inv}
({}^1 f_{ij})dx^i dx^j=d\chi^2+\sin^2\chi (d\theta^2+\sin^2\theta d\varphi^2).
\end{equation}
The eigenfunctions of the Laplace -- Beltrami operator form a basis of unitary representations of the group of
isometries of the three-dimensional space $\Sigma$ (\ref{inv}) of constant unit curvature.
In particular, the eigenfunctions on a sphere $S^3$ are the following \cite{GMM}:
\begin{equation}\label{eigenfunctionssphere}
Y_{\lambda l m} (\chi, \theta, \varphi)=\frac{1}{\sqrt{\sin \chi}}
\sqrt{\frac{\lambda (\lambda+l)!}{(\lambda-l+1)!}}
P_{\lambda-1/2}^{-l-1/2}(\cos \chi)Y_{lm}(\theta, \varphi).
\end{equation}
Here, $P_\mu^\nu (z)$ are attached Legendre functions, $Y_{lm}(\theta, \varphi)$ are spherical functions,
indices run the following values
$$\lambda=1,2,\ldots;\qquad l=0,1,\ldots,\lambda-1; \qquad m=-l,-l+1,\ldots, l.$$
There is a condition of orthogonality and normalization for functions  (\ref{eigenfunctionssphere})
\begin{equation}\label{ort}
\int\limits_0^\pi d\chi\sin^2\chi\int\limits_0^\pi
d\theta\sin\theta\int\limits_0^{2\pi} d\varphi\,Y_{\lambda l m}^{*}(\chi, \theta, \varphi)
Y_{\lambda' l' m'}(\chi, \theta, \varphi)=\delta_{\lambda\lambda'}\delta_{ll'}\delta_{mm'}.
\end{equation}
In a hyperbolic case with negative curvature ${\cal K}=-1$ the eigenfunctions are following \cite{GMM}:
\begin{equation}\label{eigenfunctionspseudosphere}
Y_{\lambda l m} (\chi, \theta, \varphi)=\frac{1}{\sqrt{\sinh \chi}}
\frac{|\Gamma(\imath\lambda+l+1)|}{|\Gamma (\imath\lambda)|}
P_{\imath\lambda-1/2}^{-l-1/2}(\cosh \chi)Y_{lm}(\theta, \varphi),
\end{equation}
where $0\le\lambda <\infty; l=0,1,2,\ldots; m=-l,-l+1,\ldots, l$.
There is a condition of orthogonality and normalization for functions  (\ref{eigenfunctionspseudosphere})
$$\int\limits_0^\Lambda d\chi\sinh^2\chi\int\limits_0^\pi d\theta\sin\theta\int\limits_0^{2\pi}
d\varphi\,Y_{\lambda l m}^{*}(\chi, \theta, \varphi)Y_{\lambda' l' m'}(\chi, \theta, \varphi)=
\delta (\lambda-\lambda')\delta_{ll'}\delta_{mm'},$$
where $\Lambda$ is some cut-off limit.

The equation on eigenvalues can be presented in the following symbolic form
\begin{equation}\label{LaplaceB}
(({}^1\bar{\rm\Delta})+k^2)Y_{\bf k}^{(s)}=0,
\end{equation}
where $-k^2$ is an eigenvalue of the Laplace -- Beltrami operator $({}^1\bar{\rm\Delta})$ on $\Sigma$.
The connection $({}^1\bar\nabla)$ is associated with the metric of invariant space (\ref{inv}).
For a flat space $({\cal K}=0)$ the eigenvectors of the equation (\ref{LaplaceB})
are flat waves as the unitary irreducible representations of the Euclidean translation group.
For a positive curvature space $({\cal K}>0)$  we have $k^2=l(l+2)$, and for a space of negative curvature
$({\cal K}<0)$ we have $k^2>1$.
The scalar contributions of the vector and the symmetric tensor fields can be expanded in terms of
\begin{eqnarray}
Y_{{\bf k},i}^{(s)}:&=&-\frac{1}{k}({}^1\bar\nabla_i) Y_{\bf k}^{(s)},\label{pr1}\\
Y_{{\bf k},{ij}}^{(s)}:&=&\frac{1}{k^2}({}^1\bar\nabla_i) ({}^1\bar\nabla_j) Y_{\bf k}^{(s)}+
\frac{1}{3}({}^1 f_{ij})Y_{\bf k}^{(s)}\label{pr2} .
\end{eqnarray}
Different modes do not couple in the linearized approximation.
So we are able to consider a contribution of a generic mode.

Let us start with considering scalar modes because of their main contribution to galaxy formation. Linear
perturbations of the four-metric in terms of the Bardeen gauge invariant potentials $\rm\Psi (t)$ and
$\rm\Phi (t)$ in conformal--Newtonian gauge are of the
\begin{equation}\label{gmn}
ds^2=-N_D^2\left(1+\rm\Psi Y_{\bf k}^{(s)}\right)^2dt^2+
a^2\left(1-\rm\Phi Y_{\bf k}^{(s)}\right)^2 ({}^1 f_{ij}) dx^i dx^j,
\end{equation}
where a summation symbol over harmonics is omit, $N_D$ is the Dirac's lapse function defined as
\begin{equation}\label{NDirac}
N\equiv N_D \left(1+\rm\Psi Y_{\bf k}^{(s)}\right).
\end{equation}
The spatial metric is presented as a sum of the Friedmann metric (\ref{Fmetric})
considered in the previous section and a perturbed part
\begin{equation}\label{Fmetricp}
(\gamma_{ij})=a^2(t)({}^1 f_{ij})-2a^2(t){\rm\Phi} Y_{\bf k}^{(s)}({}^1 f_{ij}).
\end{equation}
The determinant of the spatial metric (\ref{Fmetricp}) in the first order of accuracy is
\begin{equation}\label{gfij}
{\det} (\gamma_{ij})=a^6(t)\left(1-6\rm\Phi Y_{\bf k}^{(s)}\right){\det} ({}^1 f_{ij}).
\end{equation}
For the high symmetric Friedmann case, the background metric (\ref{back}) coincides with the conformal one:
$f_{ij}=\tilde\gamma_{ij}$
because of
\begin{equation}\label{gammafij}
\left(\frac{\gamma}{f}\right)^{1/3}=\left(\frac{a(t)}{a_0}\right)^2
\left(1-2\rm\Phi Y_{\bf k}^{(s)}\right)=e^{-2D(t)}\left(1-2\rm\Phi Y_{\bf k}^{(s)}\right).
\end{equation}
One obtains from relations (\ref{Fmetricp}), (\ref{gammafij}) the connection between components of perturbed
metric tensor and the background one
\begin{equation}\label{gammaijfij}
(\gamma_{ij})=e^{-2D(t)}\left(1-2\rm\Phi Y_{\bf k}^{(s)}\right)(f_{ij}),\qquad
(\gamma^{ij})=e^{2D(t)}\left(1+2\rm\Phi Y_{\bf k}^{(s)}\right)(f^{ij}).
\end{equation}

Now, we calculate components of the tensor of extrinsic curvature
\begin{equation}\label{Kzeroth}
K_{ij}=\frac{1}{N_D}\left[\frac{d D}{dt}-\left(({\rm\Psi}+2{\rm\Phi})\frac{d D}{dt}-
\frac{d\rm\Phi}{dt}\right)Y_{\bf k}^{(s)}\right]\gamma_{ij},
\end{equation}
and its trace
\begin{equation}
K=\gamma^{ij}K_{ij}=\frac{3}{N_D}\left[\frac{d D}{dt}-\left(({\rm\Psi}+2{\rm\Phi})
\frac{d D}{dt}-\frac{d\rm\Phi}{dt}\right)
Y_{\bf k}^{(s)}\right].
\end{equation}
Then we obtain
\begin{equation}
\nonumber
(K_{ij}K^{ij}-K^2)=-\frac{6}{N_D^2}\left[
\left(\frac{d D}{dt}\right)^2-2\left(\frac{d D}{dt}\right)
\left(({\rm\Psi}+2{\rm\Phi})\frac{d D}{dt}-\frac{d\rm\Phi}{dt}\right)Y_{\bf k}^{(s)}
\right].
\end{equation}
From the relation between metric tensors (\ref{gammaijfij}) one obtains the relation between the
corresponding Jacobians
$$\sqrt\gamma=e^{-3D(t)}\left(1-3\rm\Phi Y_{\bf k}^{(s)}\right)\sqrt{f}.$$
The calculation gives
\begin{eqnarray}
\pi_D\frac{d D}{dt}=4\sqrt{\gamma}K\frac{d D}{dt}&=&\frac{12}{N_D}e^{-3D}\sqrt{f}
\left(\frac{d D}{dt}\right)^2-\nonumber\\
&-&\frac{12}{N_D}e^{-3D}\sqrt{f}\frac{d D}{dt}
\left(({\rm\Psi}+5{\rm\Phi})\frac{d D}{dt}-\frac{d\rm\Phi}{dt}\right)
Y_{\bf k}^{(s)}.\nonumber
\end{eqnarray}
We need to the perturbation of the Ricci scalar additionally.
According to the Palatini identity \cite{Lifshitz} from the differential geometry, a variation of the Ricci
tensor is set by the formula
\begin{equation}\label{Palatini}
\delta R_{ij}=\frac{1}{2}\left(\nabla_n\nabla_j\delta\gamma_i^n+
\nabla_n\nabla_i\delta \gamma_j^n-\nabla_j\nabla_i\delta \gamma_n^n-
{\rm\Delta}\delta\gamma_{ij}\right).
\end{equation}
Here the Levi--Civita connection $\nabla_i$ is associated with the Friedmann metric $\gamma_{ij}$
(\ref{Fmetric}).
The metric variations are expressed through harmonics (\ref{Fmetricp})
\begin{equation}\label{deltagammaij}
\delta\gamma_{ij} = -2a^2(t){\rm\Phi} Y_{\bf k}^{(s)}({}^1 f_{ij}),
\end{equation}
the indices of the metric variations are moved up with respect to the background metric
\begin{equation}
\delta\gamma_i^n=\gamma^{nj}\delta\gamma_{ji}=-2{\rm\Phi} Y_{\bf k}^{(s)}\delta_i^n,\qquad
\delta\gamma_n^n=-6{\rm\Phi} Y_{\bf k}^{(s)}.
\end{equation}
Substituting them into the (\ref{Palatini}), one obtains
$$
\delta R_{ij}={\rm\Phi}\left[(\nabla_j\nabla_i-\nabla_i\nabla_j)+\nabla_j\nabla_i+
\gamma_{ij}{\rm\Delta}\right]Y_{\bf k}^{(s)}.
$$
Using the Laplace -- Beltrami equation (\ref{LaplaceB}), commutativity of the covariant differentiation
operators, and properties of harmonics (\ref{pr1}), (\ref{pr2}), one gets
\begin{equation}
\delta R_{ij}={\rm\Phi} k^2\left[Y_{{\bf k},{i j}}^{(s)}-\frac{4}{3} \left({}^1 f_{ij}\right)
Y_{\bf k}^{(s)}\right].
\end{equation}
Remark the connection between the operators, used above
$\left(\rm\Delta\right)=\left({}^1\bar{\rm\Delta}\right)/{a^2}.$
For getting the variation of the Ricci scalar we make summation
$$\delta R=\gamma^{ij}\delta R_{ij}=-\frac{4}{a^2}{\rm\Phi} k^2 Y_{\bf k}^{(s)},$$
where we used the property of traceless of the tensor
$Y_{{\bf k},i}^{i(s)}=0.$
Finally, we yield
\begin{equation}\label{Ricciscalar}
\sqrt\gamma R=\frac{6{\cal K}}{a_0^2}\sqrt{f}e^{-D(t)}-
\frac{4}{a_0^2}\sqrt{f}e^{-D(t)}{\rm\Phi} k^2 Y_{\bf k}^{(s)}.
\end{equation}

Let us consider the first order corrections to terms of the action (\ref{Wh}),
taking into account the connection between lapse functions (\ref{NDirac}),
\begin{eqnarray}
&&N\sqrt\gamma\left(K_{ij}K^{ij}-K^2\right)=-\frac{6}{N_D}e^{-3D}\sqrt{f}\left(\frac{d D}{dt}\right)^2-
\label{piDfirst}\\
&-&\frac{6}{N_D}e^{-3D}\sqrt{f}\left[-({\rm\Psi} + 7{\rm\Phi})\left(\frac{d D}{dt}\right)^2+
2\left(\frac{d D}{dt}\right)
\left(\frac{d{\rm\Phi}}{dt}\right)\right]Y_{\bf k}^{(s)}.\nonumber
\end{eqnarray}
The curvature term after the correction is the following
\begin{equation}
N\sqrt\gamma R=\frac{2}{a_0^2}N_D\sqrt{f}e^{-D}\left(3{\cal K}+
(3{\cal K}{\rm\Psi}-2k^2{\rm\Phi})Y_{\bf k}^{(s)}\right).
\end{equation}
The matter term with the correction has a form
\begin{equation}
N\sqrt\gamma T_{\bot\bot}=N_De^{-3D}\sqrt{f}\rho\left(1+(\rm\Psi-3\rm\Phi)Y_{\bf k}^{(s)}\right).
\end{equation}
The Hamiltonian constraint, multiplied to the lapse function, is the following
\begin{eqnarray}
N{\cal H}_\bot&=&-\frac{6}{N_D}e^{-3D}\sqrt{f}\left(\frac{d D}{dt}\right)^2-
\frac{6{\cal K}}{a_0^2}N_De^{-D}\sqrt{f}+N_D e^{-3D}\sqrt{f}\rho+\nonumber\\
&+&\frac{6}{N_D}e^{-3D}\sqrt{f}\left[({\rm\Psi}+7{\rm\Phi})\left(\frac{d D}{dt}\right)^2-
2\left(\frac{d D}{dt}\right)
\left(\frac{d{\rm\Phi}}{dt}\right)\right]Y_{\bf k}^{(s)}-\nonumber\\
&-&\frac{2}{a_0^2}N_De^{-D}\sqrt{f}(3{\cal K}{\rm\Psi}-2k^2{\rm\Phi})Y_{\bf k}^{(s)}+
N_De^{-3D}\sqrt{f}\rho ({\rm\Psi}-3{\rm\Phi}))Y_{\bf k}^{(s)}.\nonumber
\end{eqnarray}
For the first order correction to the Lagrangian we get the following expression
\begin{eqnarray}
&&\left(\pi_D\frac{d D}{dt}+N{\cal H}_\bot\right)^{(1)}=-\frac{6}{N_D}e^{-3D}\sqrt{f}({\rm\Psi}+3{\rm\Phi})
\left(\frac{d D}{dt}\right)^2 Y_{\bf k}^{(s)}-\nonumber\\
&-&\frac{2}{a_0^2}N_De^{-D}\sqrt{f}(3{\cal K}{\rm\Psi}-2k^2{\rm\Phi})Y_{\bf k}^{(s)}+
N_D e^{-3D}\sqrt{f}\rho({\rm\Psi}-3{\rm\Phi})Y_{\bf k}^{(s)}.\label{Lagf}
\end{eqnarray}
Then, taking into account $\sqrt{f}=a_0^3\sqrt{({}^1f)}$ and (\ref{Sigma0}),
after integrating over the space, we obtain the first order correction to the action (\ref{Wh})
\begin{eqnarray}\label{WH1}
W^{(1)}&=&6 V_0
\int\limits_{t_I}^{t_0}\, dt \frac{e^{-3D}}{N_D}({\rm\Psi}+3{\rm\Phi})\left(\frac{d D}{dt}\right)^2 Y_0^{(s)}+\\
&+&V_0\int\limits_{t_I}^{t_0}\, dt N_D\left[\frac{6 e^{-D}}{a_0^2}{\cal K}{\rm\Psi}-
e^{-3D}\rho({\rm\Psi}-3{\rm\Phi})\right]Y^{(s)}_0.\nonumber
\end{eqnarray}
In force of orthogonality of the basis functions (\ref{ort}), there are left only
zero harmonics $Y^{(s)}_0$ in (\ref{WH1})
$$\int\limits_{\Sigma}\, d^3x\sqrt{({}^1 f)}Y_{\bf k}^{(s)}=Y^{(s)}_0.$$
In the expression (\ref{WH1}), there was utilized a condition
$$\int\limits_{\Sigma}\, d^3x\sqrt{({}^1 f)}k^2 Y_{\bf k}^{(s)}=0.$$

Adding (\ref{Whint}) and (\ref{WH1}) $W=W^{(0)}+W^{(1)}$, one yields
\begin{eqnarray}
W=&-&6V_0\int\limits_{t_I}^{t_0} dt\frac{e^{-3D}}{N_D}\left(\frac{d D}{dt}\right)^2
\left[1-({\rm\Psi}+3{\rm\Phi}) Y_0^{(s)}\right]+\nonumber\\
&+&\frac{6V_0}{a_0^2}{\cal K}\int\limits_{t_I}^{t_0} dt{N_D}e^{-D}\left[1+
{\rm\Psi} Y_0^{(s)}\right]-\nonumber\\
&-&V_0\rho\int\limits_{t_I}^{t_0} dt{N_D}e^{-3D}\left[1+({\rm\Psi}-3{\rm\Phi})Y_0^{(s)}\right].\label{WHf}
\end{eqnarray}
Introducing the momentum $p_D$ by the rule
$$
p_D=\frac{\delta W}{\delta\dot D}=-\frac{12V_0}{N_D}e^{-3D}\left(\frac{d D}{dt}\right)
\left[1-({\rm\Psi}+3{\rm\Phi})Y_0^{(s)}\right],$$
we rewrite the action (\ref{WHf}) in the Hamiltonian form
\begin{equation}\label{WHHam}
W=\int\limits_{t_I}^{t_0}\, dt\left[p_D\frac{d D}{dt}+N_D {H}_\bot\right],
\end{equation}
where the Hamiltonian constraint is a following
\begin{eqnarray}
&&{H}_\bot=\frac{6V_0}{N_D^2}e^{-3D}\left(\frac{d D}{dt}\right)^2
\left[1-({\rm\Psi}+3{\rm\Phi}) Y_0^{(s)}\right]+\\
&+&\frac{3V_0}{a_0^2}{\cal K} e^{-D}\left[1+{\rm\Psi} Y_0^{(s)}\right]-
V_0\rho e^{-3D}\left[1+({\rm\Psi}-3{\rm\Phi}) Y_0^{(s)}\right].\nonumber
\end{eqnarray}
Thus we obtain the corrected Friedmann equation (\ref{Hubble0})
\begin{equation}\label{Hubblemod}
{H^2\equiv\left(\frac{d D}{dt}\right)^2=\frac{1}{6}N_D^2\rho_{mod},}
\end{equation}
where additions to the density and to the term of curvature are appeal
\begin{equation}\label{rhomod}
{\rho_{mod}=\rho \left[1+2{\rm\Psi}Y_0^{(s)}\right]+
\left[\rho_{curv}-\frac{6{\cal K}}{a^2}(2{\rm\Psi}+3{\rm\Phi}) Y_0^{(s)}\right].}
\end{equation}
To make recalculation in the conformal variables, one should replace the density in
(\ref{rhomod}) according to the conformal transformation:
$$\tilde\rho=\exp{(-4D})\rho.$$
Thus, considering the first perturbation correction of the metric, we obtain the corrected energy of the
Universe and the modified Friedmann equation.
Note, that only zero harmonics of the perturbations considered are included in the (\ref{rhomod}).

$\bullet$
Vector harmonic functions obey to the equation
$$(({}^1\bar{\rm\Delta})+k^2)Y_{{\bf k},i}^{(v)}=0.$$
They are divergent-free
$$({}^1 \bar\nabla_i) Y_{\bf k}^{(v)i}=0.$$
Tensor perturbations are constructed as
$$Y_{{\bf k},ij}^{(v)}\equiv -\frac{1}{2k}\left(({}^1\bar\nabla_i)
Y_{{\bf k},j}^{(v)}+({}^1\bar\nabla_j) Y_{{\bf k},i}^{(v)}\right).$$
Linear perturbations of the metric are taking of the form
\begin{equation}\label{gmnvec}
(g_{\mu\nu})=a^2(t)
\left(
\begin{array}{cc}
    -1&0\\
0&({}^1f_{ij})
\end{array}
\right)+a^2(t)
\left(
\begin{array}{cc}
    0&-B Y_{{\bf k},i}^{(v)}\\
-B Y_{{\bf k},j}^{(v)}&2H_T Y_{{\bf k},{i j}}^{(v)}
\end{array}
\right)
\end{equation}
with some arbitrary functions of time $B(t)$, $H_T(t)$.
The determinant of the spatial metric is
\begin{equation}\label{detvector}
{\det}(\gamma_{ij})=a^6(t).
\end{equation}
The traceless property of $Y_{{\bf k},i}^{(v)i}=0$ under the calculation of (\ref{detvector}) was used.
The inverse metric tensor is
\begin{equation}\label{gmnvecinverse}
(g^{\mu\nu})=\frac{1}{a^2(t)}
\left(
\begin{array}{cc}
    -1&0\\
0&({}^1 f^{ij})
\end{array}
\right)+\frac{1}{a^2(t)}
\left(
\begin{array}{cc}
    0&-B Y_{\bf k}^{(v)i}\\
-B Y_{\bf k}^{(v)j}&-2H_T Y_{\bf k}^{(v)ij}
\end{array}
\right).
\end{equation}

$\bullet$
Tensor harmonic functions obey to the equation on eigenvalues:
$$(({}^1\bar{\rm\Delta})+k^2)Y_{{\bf k},ij}^{(t)}=0.$$
They have properties of traceless
$$Y_{{\bf k},i}^{(t)i}=0,$$
and divergence-free
$$({}^1\bar\nabla_j) Y_{\bf k}^{(t)ij}=0.$$

Perturbations of the metric are presented as
\begin{equation}\label{gmnt}
(g_{\mu\nu})=a^2(t)
\left(
\begin{array}{cc}
    -1&0\\
0&({}^1 f_{ij})
\end{array}
\right)+a^2(t)
\left(
\begin{array}{cc}
    0&0\\
0&2H_T Y_{{\bf k},i j}^{(t)}
\end{array}
\right).
\end{equation}
The determinant of the spatial metric is
\begin{equation}\label{dettensor}
\det (\gamma_{ij})=a^6(t),
\end{equation}
in force of the property $Y_{{\bf k},i}^{(t)i}=0$.
The inverse metric tensor is
\begin{equation}\label{gmnteninverse}
(g^{\mu\nu})=\frac{1}{a^2(t)}
\left(
\begin{array}{cc}
    -1&0\\
0&({}^1f^{ij})
\end{array}
\right)+\frac{1}{a^2(t)}
\left(
\begin{array}{cc}
    0&0\\
0&-2H_T Y_{\bf k}^{(t)ij}
\end{array}
\right).
\end{equation}

We see that in force of (\ref{detvector}), (\ref{dettensor}),
the vector and tensor perturbations in linear approximation do not influence the intrinsic time $D$.





\end{document}